\documentclass[reprint, nofootinbib,preprintnumbers,amsmath,amssymb]{revtex4}
\usepackage{varioref,exscale,latexsym,amsmath,amssymb}
\usepackage{graphicx}

\usepackage{tikz}
\usepackage{slashed}
\usepackage{dcolumn}
\usepackage{bm}
\usepackage{hyperref}
\usepackage{color}
\usepackage{xcolor}

\usepackage{slashed}
\usepackage{dcolumn}
\usepackage{bm}
\usepackage{hyperref}
\usepackage{color}
\usepackage{xcolor}
\usepackage{booktabs}
\usepackage[normalem]{ulem}
\usepackage{braket}

\newcommand{\beq}{\begin{equation}}
\newcommand{\eeq}{\end{equation}}
\newcommand{\bea}{\begin{eqnarray}}
\newcommand{\eea}{\end{eqnarray}}

\newcommand{\Tr}{\mathop{\rm Tr}}

\def\lsi{\raise0.3ex\hbox{$<$\kern-0.75em\raise-1.1ex\hbox{$\sim$}}}
\def\gsi{\raise0.3ex\hbox{$>$\kern-0.75em\raise-1.1ex\hbox{$\sim$}}}

\def\beq{\begin{equation}}

\def\eeq{\end{equation}}

\def\beqa{\begin{eqnarray}}

\def\eeqa{\end{eqnarray}}

\begin{document}

\title{{\bf Do $\Lambda_{CC}$ and $G$ run?}}

\medskip\

\medskip\

\author{John F. Donoghue}
\email{donoghue@physics.umass.edu}
\affiliation{
Department of Physics,
University of Massachusetts,
Amherst, MA  01003, USA\\}

\begin{abstract}
\center{No}
\end{abstract}
\maketitle

\section{Introduction}

When renormalizing coupling constants, one often finds a scale dependence such that the coupling constant measured at one energy scale will have large corrections when used  in reactions at a different scale. The use of a running coupling will capture this scale dependence and will describe the correct coupling constant to be used at all energy scales. In the gravitational interactions there are often attempts to make $G$ and $\Lambda_{CC}$ into running couplings and to use the running versions in phenomenology\footnote{A representative but incomplete set of examples in favor of running behavior is \cite{Shapiro:2009dh, Babic:2001vv, Mavromatos:2023jsb, Bonanno:2000ep, Elizalde:1994av, Shapiro:2004ch, Hamber:2005dw}, and arguments against running are found in \cite{Hamber:2013rb, Barvinsky:2003kg, Barvinsky:2023exr, Mottola:2022tcn,  Anber:2011ut, Anber:2010uj, Donoghue:2019clrj}. }. That is the topic of this short note.

The one word abstract used above is meant to capture the essential point of this discussion. By itself it does not capture the full nuances of the discussion, which is the goal of the rest of this paper. One can always make theoretical constructs which look like running couplings, and these may have some utility in certain contexts. But the question of the title, and the answer of the abstract, is meant to refer to running in the same sense that we use running couplings such as the QCD coupling in the Standard Model - as running couplings describing the changes with the energy scale of physical amplitudes induced by quantum corrections.  As usual, energy and distance scales are inversely related, so this would apply to distance dependent couplings in gravity. 

\section{Varieties of running}

There are various techniques used when describing running coupling constants, and in the most familiar settings these give the same results. This similarity leads to a lack of focus on the physics when dealing with somewhat non-standard settings. In this section, I give a review of these techniques with a focus on their differences, using a conventional example. I will distinguish {\em physical running} which refers to running of physical amplitudes with the energy scale of the problem, {\em cutoff running} when using dimensional cutoffs for regularization and {\em $\mu$ running} when using dimensional regularization.  I will also use the Passarino-Veltman reduction to highlight the differences of scalar tadpole and scalar bubble diagrams,  and will describe non-local effective actions as the way to describe running parameters. 

\subsection{Example - chiral perturbation theory}\label{CPT}

A sample theory which is useful to use as an example is chiral perturbation theory - the low energy limit of QCD describing the interactions of spin-zero particles called pions. It is a non-linear effective field theory which has many similarities to general relativity. It has also been studied theoretically and experimentally for decades, so that we know exactly how it works.  The theory satisfies a chiral symmetry, which need not be described here (see for example \cite{Gasser:1983yg, Gasser:1984gg, Donoghue:2022wrw}) but leads to an effective Lagrangian which can be described in a derivative expansion. For massless pions, this has the form
\bea
{\cal L} &=& {\cal L}_2 + {\cal L}_4 + {\cal L}_6 + {\cal L}_8 + 
\ldots  \nonumber  \\
&=& {F^2\over 4} \Tr \left(\partial_\mu U \partial^\mu
U^\dagger\right) +
\ell_1 [\Tr \left(\partial_\mu U \partial^\mu 
U^\dagger\right)]^2   \nonumber \\
&~&+ \ell_2 \Tr \left(\partial_\mu U \partial_\nu 
U^\dagger\right) \cdot \Tr 
\left(\partial^\mu U \partial^\nu 
U^\dagger\right) + \ldots \ \ .
\eea
Here the pion fields are described by an exponential function
\beq
U(x) = \exp\left(i\frac{\tau\cdot \phi}{F} \right) \  \ .
\eeq
The constant $F$ is called the pion decay constant, $F\sim 93$ MeV. Because it multiplies the two derivative Lagrangian, it plays the role that the Planck mass does in general relativity with $F^2$ being roughly identified with $1/G\sim M_P^2$.  The coefficients $\ell_1,~\ell_2$ multiply terms of order 4 derivatives, in analogy with the curvature squared terms of general relativity. This effective field theory of pions behaves similarly to the effective field theory of general relativity \cite{Donoghue:1994dn, Donoghue:2022eay}.

Physical processes are independent of the method regularizing the divergences of the theory. The functional form can depend on the ways that we chose to define and measure the coupling constants - i.e. the renormalization scheme - but the content of the resulting formulas are the same. For example, using a particular renormalization scheme defining the couplings at $s=t=u=\mu_R^2$ all regularization methods would yield the amplitude for $\pi^+ +\pi^0 \to \pi^+ +\pi^0$ as \cite{Donoghue:2022wrw}
\bea\label{amplitude}
{\cal M} &=& {t\over F^2} +\left[8\ell^r_1(\mu_R)+2\ell_2^r(\mu_R)\right]{t^2\over F^4} \nonumber \\
&+&2\ell_2^r(\mu_R)\frac{s(s-u) +u(u-s)}{F^4} \\
&-& {1\over 96\pi^2 F^4}\left[3t^2 \ln {-t\over \mu_R^2}+s(s-u)\ln {-s\over \mu_R^2}+u(u-s)\ln {-u\over \mu_R^2}\right]  \ \ , \nonumber 
\eea
with $s, ~t,~u$ being the usual Mandelstam variables. However, there will be differences on how different regularization methods arrive at this result, and I will describe these differences in the following subsections. 

The scattering amplitude contains the visible results of a power counting theorem of Weinberg \cite{Weinberg:1978kz}.  Higher loops renormalize higher order operators in the effective Lagrangian. In particular one loop corrections renormalize the operators of order $(\rm energy)^4$ but not the leading term of order $E^2$. We see this in the amplitude because the divergences and the logs appear at order $E^4$. The gravitational effective field theory has a similar power counting theorem, and we will see that it is relevant to the topic of this paper. Weinberg's theorem uses dimensional regularization in its proof, because that method does not introduce any powers of dimensionful parameters. We will see that cutoff regularization violates this. But because all physical quantities are independent of regularization method, powers of the cutoff must be absorbed into renormalized constants without leaving behind any physical remnant in amplitudes.

\subsection{Physical running}

In particle physics, the use of running couplings refers to the dependence of physical processes on the energy scale. When a reaction is observed at some energy scale, one uses a running coupling appropriate for that energy. This captures the quantum processes relevant for that energy and avoids the appearance of large logarithms which could spoil perturbation theory. 

This use is visible in the amplitude of Eq. \ref{amplitude}. The coefficients $\ell_1,~\ell_2$ could have been measured to be particular values near some renormalization scale $\mu_R$, but when applied at energies which are far different one should use different values determined by the renormalization group in order to avoid the appearance of large logarithms. The physical invariance of the scattering amplitude leads to the beta functions
\bea
\beta_{\ell_1} = \mu_R \frac{\partial}{\partial \mu_R} \ell_1 =  -\frac1{96\pi^2}\nonumber \\
\beta_{\ell_2} = \mu_R \frac{\partial}{\partial \mu_R} \ell_2 =-\frac1{192\pi^2}
\eea
These are typical running couplings. The running of yet higher order chiral logarithms is also well understood in chiral perturbation theory \cite{Buchler:2003vw}.

However, one can also see that $F$ does not run. The portion of the amplitude which is governed by $F$ has no energy dependence. If we measure $F$ at one energy scale, the same value can be used at all energies. The beta function for the leading order coupling is 
\beq
\beta_F = \mu_R \frac{\partial}{\partial \mu_R} F = 0
\eeq

\subsection{Cutoff running}\label{cutoffrunning}

If one uses an energy cutoff in describing quantum corrections, the results for any observable will depend on the cutoff for any finite value of the cutoff. One can consider lattice calculations within QCD as an archetype for this. The lattice spacing $a$ provides a UV cutoff of order $\Lambda\sim 1/a$. The calculation of $F$ or any other parameter will depend on $a$ when calculated at finite values of $a$.  The physical value is obtained by taking the continuum limit $a\to 0$ ($\Lambda \to \infty$) . Here the $a$ dependence of the calculation does not imply that $F$ is a running coupling. It simply reflects that  that the calculation is incomplete, and there are quantum corrections from energy scales beyond $\Lambda$ which have not yet been included in the calculation. When performing a complete calculation using lattice QCD, one can actually account for these contributions by extrapolating to the continuum. The result is the physical coupling. 

When using a cutoff as a regulator in perturbation theory, the logic is similar. We calculate the physics that we know up to some scale $\Lambda$ and recognize that there is potentially unknown physics beyond that scale. We would calculate it if we could. But the beauty of the renormalization program is that all that physics, the cutoff and all that lies beyond it, disappears when we use the measured parameters, which obviously include all the physics which is found in Nature.

\begin{figure}[htb]
\begin{center}
\includegraphics[height=30mm,width=80mm]{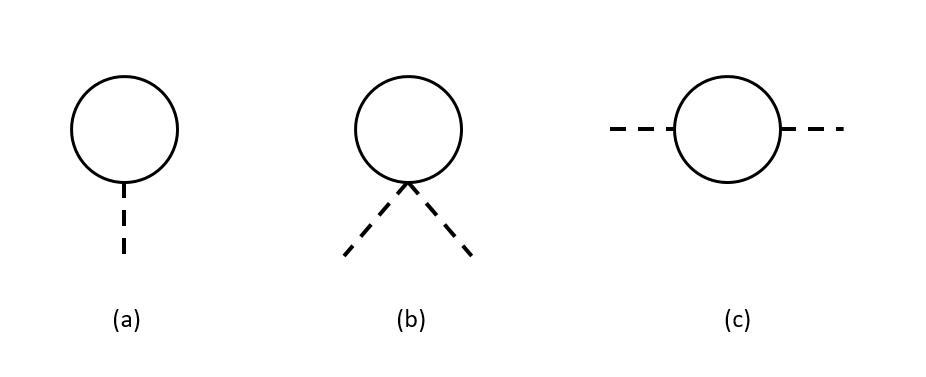}
\caption{Examples of  tadpole diagrams (a) and (b) and the bubble diagram (c). The dashed lines can represent various external states but the key feature is that external momentum flows through the bubble diagram but not through the tadpole diagrams. }
\label{diagrams}
\end{center}
\end{figure}

The use of a cutoff will lead to renormalization of the parameter $F$ which is quadratic in the cutoff $\Lambda$. This can be addressed using either the normalization of the propagator or the axial coupling. In both cases, the relevant diagram is a tadpole diagram, for example as in  Figure 1 a,b. The loop integral is 
\beq\label{tadpole}
-i \int \frac{d^4p}{(2\pi)^4} \frac1{p^2+i\epsilon} = \frac1{16\pi^2} \Lambda^2
\eeq
as it is quadratically divergent. Including the appropriate numerical factors leads to a one-loop correction to $F$
\beq
\delta F = \frac{\Lambda^2}{16\pi^2F}   \ \ .
\eeq
If we apply usual methods following the divergences of a parameter this would lead to the conclusion that $F$ runs quadratically with the cutoff
\beq
\beta_F = \Lambda \frac{\partial}{\partial \Lambda} F = \frac1{8\pi^2}\Lambda^2  \ \ .
\eeq 
However, this loop integral is not sensitive to any of the energy scales of any reaction. It is just a constant and is absorbed into the physical value of $F$ upon renormalization\footnote{If we had included a mass for the pion, there would in general be a sub-leading logarithmic dependence on the cutoff also. This will also disappear into the renormalized parameter, similar to the discussion related to $\mu$-running in the next subsection.}
\beq
F_{ren} = F_{bare} + \delta F \ \ .
\eeq

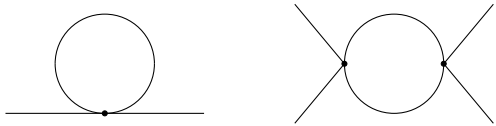
\begin{figure}[ht]\label{scatteringloops}
\begin{center}
\begin{tikzpicture}
[>=stealth,scale=0.66,baseline=-0.1cm]
\draw (-1,0) arc (180:0:1cm);
\draw (-1,0) arc (180:360:1cm);
\draw (2,-1) node[anchor=west] {} -- (0,-1);
\draw (0,-1) -- (-2,-1) node[anchor=east] {};
\draw[fill] (0,-1) circle [radius=0.05];
\end{tikzpicture}
\qquad
\begin{tikzpicture}
[>=stealth,scale=0.66,baseline=-0.1cm]
\draw (-1,0) arc (180:0:1cm);
\draw (-1,0) arc (180:360:1cm);
\draw (2,1.2) node[anchor=west] {} -- (1,0);
\draw (2,-1.2) node[anchor=west] {} -- (1,0);
\draw (-1,0) -- (-2,1.2) node[anchor=east] {};
\draw (-1,0) -- (-2,-1.2) node[anchor=east] {};
\draw[fill] (-1,0) circle [radius=0.05];
\draw[fill] (1,0) circle [radius=0.05];
\end{tikzpicture}
\caption{The diagrams 
giving corrections to the two- and four-point functions.}
\label{fig.chiral}
\end{center}
\end{figure}

When calculating the scattering amplitude at one loop, we need the wavefunction renormalization (i.e. $F$) and the scattering correction. There will be both tadpole and bubble diagrams, as in Fig. 2. By using the Passarino-Veltman reduction technique, to be reviewed below, these can be reduced to scalar tadpole and scalar bubble diagrams with no momentum factors in the numerators. The chiral symmetry ensures that the renormalization of $F$ in the scattering amplitude will be the same as was calculated above. - it will again be a tadpole with no residual energy dependence.   There will also be the scalar bubble diagram
\bea
I_B(q) &= & -i\int \frac{d^4p}{(2\pi)^4} \frac1{[p^2+i\epsilon][(p-q)^2+i\epsilon]} \nonumber \\
&=& \frac1{16\pi^2}\log \frac{\Lambda^2}{-q^2}
\eea
This contributes to the amplitude at order $(\rm Energy)^4$ as required by the Weinberg power counting theorem.  The bubble diagram is logarithmically divergent and also depends logarithmically on the energy scale. The divergences will be absorbed into the renormalized parameters $\ell_1,~\ell_2$. 
\beq
\ell_i(\mu_R) = \ell_i + \delta \ell_1(s=t=u=\mu_R^2)
\eeq
To measure these parameters one must use some renormalization scale $\mu_R$ and a renormalization scheme. This leads to the physical scattering amplitude of Eq. \ref{amplitude}

For these parameters, one could also obtain the correct beta function by following the dependence of the quantum correction on the cutoff,
\beq
\beta_i = \Lambda \frac{\partial}{\partial \Lambda} \delta \ell_i    \ \ .
\eeq
In this case this is clear since there are no other dimensional parameters in the theory and the logarithm must depend on $\log (\Lambda^2/q^2)$. 

In theories with extra factors of masses, following $\log \Lambda$ does not always give the correct beta function because of factors of $\log (\Lambda^2/m^2)$ which do not depend on the energy scale in the amplitude.  The general structure will be of the form
\beq
{\cal M} \sim g^2 \left[1 + a  \log  \frac{\Lambda^2}{m^2} +b \log \frac{\Lambda^2}{q^2}+... \right]
\eeq
with $a, ~b$ being constants. Upon renormalization of the coupling, the $\log \Lambda^2/m^2$ factors disappear into the renormalized coupling, and do not lead to physical running. After renormalization, the $\log \Lambda^2/q^2$ terms turn into $\log \mu_R^2/q^2$ physical effects, where $\mu_R$ is related to the renormalization scale at which the renormalized coupling is measured.

\subsection{$\mu$ running}

Another common technique is define the beta functions by following the parameter $\mu$ that occurs in dimensional regularization, i.e. $\mu \partial /\partial \mu$.  This always works in massless theories, because in dimensional regularization the massless tadpole diagram vanishes 
\beq
I_T =0
\eeq
and in the bubble diagram the logarithmic dependence on $\mu$  is always related to the energy dependence
\beq
I_B (q) =\frac1{16\pi^2}\left[  \frac1{\bar{\epsilon }}+ \log \frac{\mu^2}{-q^2} \right]
\eeq
with $1/\bar{\epsilon}  =2/(4-d) +\log 4\pi -\gamma$. The fact that the tadpole diagram vanishes tells us that the parameter $F$ is not renormalized at one loop. It is therefore obvious that it does not run.  In the scattering amplitude the quantum corrections will all be at the order $E^4$, and it is not hard to see that for the coefficients $\ell_i$ tracing the $\mu$ dependence 
\beq
\beta_i = \mu \frac{\partial}{\partial \mu} \delta \ell_i    
\eeq
will also give the correct beta function.

However, a point which I wish to emphasize here is that when masses are involved,  the $\mu$ dependence does not always indicate running. In the real world, the chiral effective field theory does involve the pion mass.  In this case there is a non-zero renormalization of F (or more precisely the pion decay constant of the real world $F_\pi = 93$MeV).  The loop correction to $F_\pi$ is again the tadpole diagram and has the value \cite{Gasser:1983yg}
\beq
\delta F_\pi =  \frac{m_\pi^2}{16\pi^2F}\left(\ell_4 ^r+ \log \frac{\mu^2}{m_\pi^2}\right)
\eeq
where $\ell_4$ is another coefficient which appear in the Lagrangian when including masses. However for our purposes the important point is that despite the fact that 
\beq
\mu \frac{\partial}{\partial \mu}\delta F_\pi \ne 0
\eeq
the decay constant is not a running parameter.  The quantum correction is absorbed into the renormalized value but there is no dependence of $F_\pi$ on any external energy scale which varies in physical reactions. There is a wealth of phenomenology to demonstrate this empirically. 

\subsection{Tadpoles and bubbles}

Passarino and Veltman \cite{Passarino:1978jh} showed that all one loop Feynman integrals can be reduced to functions of the scalar tadpole, bubble, triangle, and box diagrams. Here scalar means that there are no momentum factors in the numerator. The only two of these scalar integrals which are divergent are the tadpole and bubble diagrams, so that these are the ones which are involved in renormalization. 

The tadpole diagram is independent of any external momentum. It cannot lead directly lead to any energy dependence of the renormalized couplings. The quadratic $\Lambda^2$ behavior of the tadpole found when using a cutoff is then never part of a physical beta function, nor are the $m^2$ factors arising from the tadpole. The scalar bubble diagram  does carry energy dependence and has a $\log q^2$ when the momentum is greater than the mass in the loop. It is generally this logarithm which leads to the physical running of couplings. 

In a theory with no mass parameters, dimensional analysis says that the logarithms come with $\log \Lambda^2/q^2$ or $\log \mu^2/q^2$, so that following the $\Lambda$ or $\mu$ dependence will correctly identify the beta functions. However, if there are masses around, there can also be factors of $\log \Lambda^2/m^2$ or $\log \mu^2/m^2$. These do not lead to physical running, and here following $\Lambda$ or $\mu$ can be misleading.

\subsection{Nonlocal effective actions}\label{nonlocal1}

A powerful way to represent the energy dependence that leads to physical running is the use of non-local effective actions. For example, the running of the 
QED coupling due to the vacuum polarization diagram with a massless field can be represented in shorthand notation by\footnote{The reader who would like to see this derived in coordinate space can find the calculation in Ref \cite{Donoghue:2015xla}. }
\beq\label{QED}
S_{eff} = \int d^4x ~- \frac14 F_{\mu\nu} \left[\frac1{e^2 (\mu)}- b \log (\Box/\mu^2) \right] F^{\mu\nu}
\eeq
where $b$ is related to the beta function of the theory. This compact notation actually hides a non-local effective acton because the $\log \Box$ represents 
a nonlocal function
\bea\label{quasilocal}
&~&\int d^4x ~ F_{\mu\nu} \log (\Box/\mu^2) F^{\mu\nu} \nonumber  \\
& \implies& \int d^4x  d^4y ~F_{\mu\nu} (x)\langle x| \log (\Box/\mu^2) |y\rangle F^{\mu\nu}(y)  \ \ .
\eea
describing the Fourier transform of $log -q^2$,
\beq\label{logBox}
\langle y|\log \Box | x \rangle = \int \frac{d^4q}{(2\pi)^4} ~e^{iq\cdot(x-y)} \log -q^2
\eeq
The fact that physics is independent of the scale $\mu$ gives the beta function $\beta = be^2$.
However it is the $\log \Box$ which tells us that matrix elements of this effective action will depend on the energy involved in the reaction. 

In the chiral theory that we have been using as an example, these nonlocal effective actions also exist. They can be found in the so-called {\em unitarity corrections} in both of the classic papers of Gasser and Leutwyler on chiral perturbation theory \cite{Gasser:1983yg, Gasser:1984gg}. The authors also separate out the tadpole corrections from the logarithms of momenta even in the presence of masses. The curious reader is referred there for the details. For the purpose of this paper, it is not needed to describe these fully here, because a related formalism for gravity and gauge theory has been developed by Barvinsky and Vilkovisky \cite{Barvinsky:1985an, Barvinsky:1990up, Vilkovisky:1992pb}, and we will apply this directly to gravity below. 

\section{The cosmological constant}

The gravitational interaction can be organized in a derivative expansion with local terms consistent with general covariance
\beq
S_{grav} = \int d^4 x \sqrt{-g} \left[    - \Lambda_{CC} +\frac{2}{\kappa^2}M R + c_1 R^2 + c_2 R_{\mu\nu}R^{\mu\nu} +...      \right]  \ \ ,
\eeq
with $\Lambda_{CC}$ being the cosmological constant and $\kappa^2 =32\pi G$. 
The lessons discussed above also apply to gravity. We will argue that while the renormalization of $\Lambda_{CC}$ and $G$ can depend on cutoffs or on the scale factor $\mu$ in dimensional regularization, these dependences do not amount to the running of these parameters in physical processes.

\subsection{One loop renormalization}
Let us start with a simple example of the renormalization of the cosmological constant by a massive scalar field, using dimensional regularization. This can be explored in the weak field limit by calculating the coupling to the gravitational field. Using $g_{\mu\nu} = \eta_{\mu\nu} + h_{\mu\nu}$, the cosmological term in the action can be expanded
\beq
\sqrt{-g}\Lambda_{CC} = \Lambda_{CC}(1+\frac12 h^\sigma_\sigma + \frac18 ( h^\sigma_\sigma )^2 -\frac14 h_{\sigma\lambda}h^{\sigma\lambda} +...)  \ \ .
\eeq
One can study the renormalization of the cosmological constant by calculating the renormalization of the one-point, two-point etc coupling in the weak field limit. It is simplest and most instructive to calculate the coupling to a single $h_{\mu\nu}$ field. This is just the tadpole loop of Fig. \ref{diagrams}a. 
One finds
\beq\label{deltaLambda}
\delta \Lambda = -\frac{m^4}{32\pi^2} \left[  \frac1{\epsilon} -\gamma +\log(4\pi) +\log \frac{\mu^2}{m^2} +\frac32 \right] \ \ . 
\eeq
This result depends on the parameter $\mu$ and exhibits what I referred to above as $\mu$-running. 
\beq
\mu \frac{\partial \delta\Lambda}{\partial \mu} =  -\frac{m^4}{16\pi^2}  \ne 0  \ \ .
\eeq 
However, this cannot depend on any external momentum scale and the quantum correction just provides a constant shift in the value of the cosmological constant. When renormalized,
\beq
\Lambda_{CC}= \Lambda_{bare} +\delta \Lambda_{CC}
\eeq
the measured value will be the same at any energy scale.  This implies that there is no physical running of the cosmological constant.

The fact that you can renormalize the cosmological constant by using the single graviton coupling - which always must be independent of the external energy scales - makes this a general result. One might worry that the second order coupling will bring in a bubble diagram, and this could bring in energy dependence. However, explicit calculation yields the same result for the renormalization. 

When using cutoff regularization in the evaluation of this integral, it is common to say that the result is of fourth order in the cutoff, i.e. $\Lambda^4$. However there is a little appreciated contribution from the path integral measure which cancels that contribution, arising from the path integral measure  \cite{Fradkin:1974df,  Donoghue:2020hoh}. There are residual effects at order $m^2\Lambda^2$ and $m^4 \log \Lambda$. However, because these can be calculated from the tadpole diagram there is no energy dependence. Like the $\mu$ dependence discussed above, the quantum correction is a constant and is absorbed into the renormalized value of the cosmological constant.

\subsection{Non-local actions}

At one loop, gravitons and matter fields produce divergences at the order of the curvature squared. These are local terms in position space, and the divergences can be absorbed into the coefficients in the local action. Accompanying these divergences come logarithmic corrections. In momentum space, these are logs of the momenta found in the bubble diagram. As described above in Sec. \ref{nonlocal1}, in coordinate space, they are generally notated as $\log \Box$, which is the nonlocal function of Eq. \ref{logBox}.   Barvinsky and Vilkovisky and collaborators have developed this program as a non-local expansion in the curvature \cite{Barvinsky:1985an, Barvinsky:1990up, Vilkovisky:1992pb}. In curved spacetime there is considerable ambiguity in the covariant form of $\log \Box$ although the effect of that ambiguity can be shifted to higher order in the curvature expansion. After renormalization, the curvature squared terms of the effective action are
\bea\label{BV}
S_{grav} &=& \int d^4 x \sqrt{-g} \left[ c_1(\mu) R^2 +d_1 R\log \left(\frac{ \Box}{\mu^2 } \right)R \right. \nonumber \\ 
 &+&\left. c_2(\mu) R_{\mu\nu}R^{\mu\nu} +d_2 R_{\mu\nu} \log \left(\frac{ \Box}{\mu^2 }\right)R^{\mu\nu} \right]  \ \ .'
\eea
The coefficients $d_i$ are known as these follow from the one loop divergences of gravity and matter. These tell us that the curvature squared couplings are running couplings in the physical sense.

Once we understand how running couplings are represented using nonlocal effective actions, we see clearly the lack of running of the cosmological constant. To my knowledge, this form of the argument was first given by Barvinsky \cite{Barvinsky:2003kg}.
The key point is that quantum corrections do not generate an isolated $\log \Box$ term in an effective action. The expression
\beq
S \stackrel{?}{=} \int d^4x \sqrt{-g} \left[\Lambda + d \log \Box\right]
\eeq
is not a possible quantum correction. The $\log \Box$ factor has nothing to act on.. Even if 
\beq
\int d^4 x \sqrt{-g(x)} d^4y \sqrt{-g(y)} \langle y| \log\Box |x\rangle
\eeq
were somehow interpreted to make sense, its matrix element would not contain a running coupling for the single $h_{\mu\nu} $ matrix element in the weak field limit.  So the argument here is that $\Lambda$ and also $G$ cannot run because we cannot have non-local actions for these parameters which are consistent with general covariance. 

There can be a related non-local Lagrangian which appears formally at the same order in the derivative expansion, but which however is distinct from the cosmological constant. This can be found as part of the bubble diagram of Fig. 1c. There is a left over energy dependent interaction, which has a different structure from the cosmological constant \cite{Donoghue:2022chi}. I have called this the non-local partner of the cosmological constant, and have calculated the form of the two graviton bubble diagram and used the methods of Barvinsky-Vilkovisky to make a covariant operator, finding
\beqa\label{partner}
{\cal L} &=&  \frac{m^4}{40\pi^2}\left[ \left(\frac1{\Box}R_{\lambda\sigma}\right)\log((\Box+m^2)/m^2)\left(\frac1{\Box}R^{\lambda\sigma} \right)-\frac18 \left(\frac1{\Box}R\right) \log((\Box+m^2)/m^2)\left(\frac1{\Box}R\right) \right] \nonumber \\
&+&  \frac{m^2}{240\pi^2} \left[R_{\lambda\sigma}\frac{1}{\Box}R^{\lambda\sigma} -\frac18R\frac{1}{\Box}R \right]  
\eeqa
with the shorthand notation
\beq
\langle x|  \log [(\Box+m^2/m^2) | y\rangle = \int \frac{d^4q}{(2\pi)^4} e^{iq\cdot(x-y)} \int_0^1 dx ~\log\left[\frac{m^2 -x(1-x)q^2}{m^2} \right]   \  \ .
\eeq
This operator is of zeroth order in the derivative expansion, like the cosmological constant. However, it starts at second order in the gravitational field and hence is distinct.

\subsection{Unimodular gauge}

For the cosmological constant there is an independent argument against running, which is unique to gravity. Due to the general covariance of the theory, it is possible to make a gauge choice \cite{Buchmuller:1988yn, Finkelstein:2000pg, Percacci:2017fsy, Salvio:2024vfl}
such that $\sqrt{-g}= 1$. In this gauge the vacuum energy $\Lambda_{CC}$ does not couple to the metric and the equations of motion doe not involve $\Lambda_{CC}$. However, the cosmological constant as we know it re-emerges in this gauge as an initial condition for a constraint.  The issue is that the equations of motion in this gauge do not satisfy the conservation equation for the stress tensor. There is the need for a constraint equation to enforce conservation. That equation requires the specification of an initial condition. Using this, one obtains the usual predictions of general relativity.

While the use of this gauge does not change general relativity, it is useful for the topic of this paper. The initial condition is just a number. It does not have the possibility to be a running parameter depending on the energy or distance scales.

\section{Newton's constant $G$}

The reasoning for the non-running of $G$ is roughly similar. Direct calculation of the contribution of a matter loop of a massive particle yields a divergence 
\beq
\delta \frac{1}{G} \sim m^2
\eeq
without any $\log q^2$. As with $\Lambda_{CC}$ this will be absorbed into the renormalized $G$ with no residual running. This is as it must be because there cannot be a non-local effective action of the form $\Gamma^\lambda_{~\alpha\beta} \log \Box \Gamma_\lambda^{~\alpha\beta} $ as this is forbiddent by general coordinate invariance\footnote{The ``nonlocal partner'' of $G$ is the term of order $m^2$ in Eq.  \ref{partner} but this has a different structure from the Einstein action.}. 

For the case of $G$ there is an additional phenomenological case against the definition of a running constant. I have argued above that the power-law dependence on a cutoff does not constitute running behavior, using example of $F$ in the chiral perturbative scattering amplitude. In general relativity, $1/G$ plays the same role that $F$ plays in the chiral case, as the coefficient of the two-derivative term in the Lagrangian. The real running behavior happens with the four-derivative terms. However, one might try to argue that the $\Lambda^2 $ dependence was in some way a proxy for the higher derivative terms that come with extra powers of the energy. Perhaps there is some phenomenological way to define a running coupling with power-law running. In the case of gravity this has been checked and found not to work  \cite{Anber:2011ut, Anber:2010uj,  Donoghue:2019clrj}, by calculating several observables and looking for common features. Part of it is numerical, in that the quantum corrections to different processes come with very different numerical factors and signs. This is because there is no reason for them to be related, as they are not the renormalization of the same object. Part of the reason is also kinematical. Higher order corrections come with different kinematic invariants, and can be of different magnitudes and signs\footnote{This can also be seen in the chiral amplitude of Eq  \ref{amplitude}.} In some cases, corrections which are positive in, say, the $s$ channel would need to be negative in the $t$ channel. There is no useful definition of power-law running parameters.

\section{Summary}

Most of the statements and techniques described here are well-known in the context of other interactions. However, quantum gravity does not have a long tradition of phenomenological applications, and so it is perhaps less clear how these techniques apply to $G$ and $\Lambda_{CC}$. Hopefully by seeing the example of a very similar effective field theory in Sec. \ref{CPT}, one can translate them to the the case of gravity.

I would like to here also continue the discussion of the use of cutoffs, which was started by discussion of lattice cutoffs in Sec. \ref{cutoffrunning}.  When using a cutoff in a theoretical calculation, there will always be a dependence on the cutoff for intermediate results. While one can use this dependence to define a beta function, and it can be useful to do so within that calculation, it is not necessarily a function related to behavior of couplings in physical amplitudes. Rather it is an indication of an incomplete calculation. The example previously was the coupling $F$ which does not run at all in amplitudes, but which would appear to have cutoff dependence when using the lattice or any other method involving a cutoff. The real world is obtained only by removing the cutoff. While such a running coupling can be useful within the context of the particular theoretical calculation, it can be disconnected from physical running. In gravitational theories, the Asymptotic Safety program \cite{Niedermaier:2006wt, Reuter:1996cp, Percacci:2017fkn, Reuter:2019byg}
uses the Functional Renormalization Group in a similar way. The quantum corrections are accounted for above some separation scale $k$, and the couplings have a $k$ dependence. The real world is obtained by integrating  over all values down to $k=0$. The $k$ dependence involves both power-law running and logarithmic running. However, neither is guaranteed to be present in physical amplitudes. There are by now several counter examples \cite{Buccio:2023lzo, Donoghue:2023yjt, Buccio:2024hys, Buccio:2024omv}. A clear takeaway is that the running with a cutoff found in such schemes is not to be naively used in phenomenological applications of physical quantities.

\section*{Acknowledgements}

This work has been partially supported by the US National Science Foundation under grants NSF-PHY-2112800 and NSF-PHY- 2412570
The material discussed here is an elaboration of some topics presented at the 64th Cracow School of Theoretical Physics (2024) \cite{Cracow} and will be published in the proceedings. I thank Michal Praszałowicz and the other organizers of the school for their hospitality and also thank Cliff Burgess, Andrei Barvinsky, Emil Mottola, Ilya Shapiro, Antonio Pereira, Alexander Monin, Kasra Kiaee, Arkady Tseytlin, Diego Buccio, Gabriel Menezes and Roberto Percacci for discussions on this topic. 

\eject


\begin{thebibliography}{99}


\bibitem{Shapiro:2009dh}
I.~L.~Shapiro and J.~Sola,
``On the possible running of the cosmological 'constant',''
Phys. Lett. B \textbf{682}, 105-113 (2009)
doi:10.1016/j.physletb.2009.10.073
[arXiv:0910.4925 [hep-th]].

\bibitem{Babic:2001vv}
A.~Babic, B.~Guberina, R.~Horvat and H.~Stefancic,
``Renormalization group running of the cosmological constant and its implication for the Higgs boson mass in the standard model,''
Phys. Rev. D \textbf{65}, 085002 (2002)
doi:10.1103/PhysRevD.65.085002
[arXiv:hep-ph/0111207 [hep-ph]].

\bibitem{Mavromatos:2023jsb}
N.~E.~Mavromatos, J.~Sol\`a Peracaula and A.~G\'omez-Valent,
``String-inspired running-vacuum cosmology, quantum corrections and the current cosmological tensions,''
[arXiv:2307.13130 [gr-qc]].

\bibitem{Bonanno:2000ep}
A.~Bonanno and M.~Reuter,
%
``Renormalization group improved black hole space-times,''
Phys. Rev. D \textbf{62}, 043008 (2000)
doi:10.1103/PhysRevD.62.043008
[arXiv:hep-th/0002196 [hep-th]].

\bibitem{Elizalde:1994av}
E.~Elizalde, S.~D.~Odintsov and I.~L.~Shapiro,
``Asymptotic regimes in quantum gravity at large distances and running Newtonian and cosmological constants,''
Class. Quant. Grav. \textbf{11}, 1607-1614 (1994)
doi:10.1088/0264-9381/11/7/004
[arXiv:hep-th/9404064 [hep-th]].
\bibitem{Shapiro:2004ch}
I.~L.~Shapiro, J.~Sola and H.~Stefancic,
``Running G and Lambda at low energies from physics at M(X): Possible cosmological and astrophysical implications,''
JCAP \textbf{01}, 012 (2005)
doi:10.1088/1475-7516/2005/01/012
[arXiv:hep-ph/0410095 [hep-ph]].

\bibitem{Hamber:2005dw}
H.~W.~Hamber and R.~M.~Williams,
``Nonlocal effective gravitational field equations and the running of Newton's G,''
Phys. Rev. D \textbf{72}, 044026 (2005)
doi:10.1103/PhysRevD.72.044026
[arXiv:hep-th/0507017 [hep-th]].


\bibitem{Hamber:2013rb}
H.~W.~Hamber and R.~Toriumi,
``Inconsistencies from a Running Cosmological Constant,''
Int. J. Mod. Phys. D \textbf{22}, no.13, 1330023 (2013)
doi:10.1142/S0218271813300231
[arXiv:1301.6259 [hep-th]].

\bibitem{Barvinsky:2003kg}
A.~O.~Barvinsky,
``Nonlocal action for long distance modifications of gravity theory,''
Phys. Lett. B \textbf{572}, 109-116 (2003)
doi:10.1016/j.physletb.2003.08.055
[arXiv:hep-th/0304229 [hep-th]].

\bibitem{Barvinsky:2023exr}
A.~O.~Barvinsky and W.~Wachowski,
``Notes on conformal anomaly, nonlocal effective action, and the metamorphosis of the running scale,''
Phys. Rev. D \textbf{108}, no.4, 045014 (2023)
doi:10.1103/PhysRevD.108.045014
[arXiv:2306.03780 [hep-th]].

\bibitem{Mottola:2022tcn}
E.~Mottola,
``The effective theory of gravity and dynamical vacuum energy,''
JHEP \textbf{11}, 037 (2022)
doi:10.1007/JHEP11(2022)037
[arXiv:2205.04703 [hep-th]].

\bibitem{Anber:2011ut}
M.~M.~Anber and J.~F.~Donoghue,
``On the running of the gravitational constant,''
Phys. Rev. D \textbf{85}, 104016 (2012)
doi:10.1103/PhysRevD.85.104016
[arXiv:1111.2875 [hep-th]].

\bibitem{Anber:2010uj}
M.~M.~Anber, J.~F.~Donoghue and M.~El-Houssieny,
``Running couplings and operator mixing in the gravitational corrections to coupling constants,''
Phys. Rev. D \textbf{83}, 124003 (2011)
doi:10.1103/PhysRevD.83.124003
[arXiv:1011.3229 [hep-th]].

\bibitem{Donoghue:2019clrj}
J.~F.~Donoghue,
``A Critique of the Asymptotic Safety Program,''
Front. in Phys. \textbf{8}, 56 (2020)
doi:10.3389/fphy.2020.00056
[arXiv:1911.02967 [hep-th]].


\bibitem{Gasser:1983yg}
J.~Gasser and H.~Leutwyler,
``Chiral Perturbation Theory to One Loop,''
Annals Phys. \textbf{158}, 142 (1984)
doi:10.1016/0003-4916(84)90242-2

\bibitem{Gasser:1984gg}
J.~Gasser and H.~Leutwyler,
``Chiral Perturbation Theory: Expansions in the Mass of the Strange Quark,''
Nucl. Phys. B \textbf{250}, 465-516 (1985)
doi:10.1016/0550-3213(85)90492-4

\bibitem{Donoghue:2022wrw}
J.~F.~Donoghue, E.~Golowich and B.~R.~Holstein,
``Dynamics of the Standard Model: Second edition,''
Cambridge University Press, 2022,
ISBN 978-1-009-29100-2, 978-1-009-29101-9, 978-1-009-29103-3
doi:10.1017/9781009291033


\bibitem{Donoghue:1994dn}
J.~F.~Donoghue,
``General relativity as an effective field theory: The leading quantum corrections,''
Phys. Rev. D \textbf{50}, 3874-3888 (1994)
doi:10.1103/PhysRevD.50.3874
[arXiv:gr-qc/9405057 [gr-qc]].


\bibitem{Donoghue:2022eay}
J.~F.~Donoghue,
``Quantum General Relativity and Effective Field Theory,'' in {\it Handbook of Quantum Gravity} ed. by Cosimo Bambi, Leonardo Modesto and Ilya Shapiro. (Springer, Singapore 2023) 
doi:10.1007/978-981-19-3079-9\_1-1
[arXiv:2211.09902 [hep-th]].

\bibitem{Weinberg:1978kz}
S.~Weinberg,
``Phenomenological Lagrangians,''
Physica A \textbf{96}, no.1-2, 327-340 (1979)
doi:10.1016/0378-4371(79)90223-1

\bibitem{Buchler:2003vw}
M.~Buchler and G.~Colangelo,
``Renormalization group equations for effective field theories,''
Eur. Phys. J. C \textbf{32}, 427-442 (2003)
doi:10.1140/epjc/s2003-01390-2
[arXiv:hep-ph/0309049 [hep-ph]].

\bibitem{Passarino:1978jh}
G.~Passarino and M.~J.~G.~Veltman,
``One Loop Corrections for e+ e- Annihilation Into mu+ mu- in the Weinberg Model,''
Nucl. Phys. B \textbf{160}, 151-207 (1979)
doi:10.1016/0550-3213(79)90234-7



\bibitem{Donoghue:2015xla}
J.~F.~Donoghue and B.~K.~El-Menoufi,
``QED trace anomaly, non-local Lagrangians and quantum Equivalence Principle violations,''
JHEP \textbf{05}, 118 (2015)
doi:10.1007/JHEP05(2015)118
[arXiv:1503.06099 [hep-th]].

\bibitem{Barvinsky:1985an}
A.~O.~Barvinsky and G.~A.~Vilkovisky,
``The Generalized Schwinger-Dewitt Technique in Gauge Theories and Quantum Gravity,''
Phys. Rept. \textbf{119}, 1-74 (1985)
doi:10.1016/0370-1573(85)90148-6

\bibitem{Barvinsky:1990up}
A.~O.~Barvinsky and G.~A.~Vilkovisky,
``Covariant perturbation theory. 2: Second order in the curvature. General algorithms,''
Nucl. Phys. B \textbf{333}, 471-511 (1990)
doi:10.1016/0550-3213(90)90047-H

\bibitem{Vilkovisky:1992pb}
G.~A.~Vilkovisky,
``Effective action in quantum gravity,''
Class. Quant. Grav. \textbf{9}, 895-903 (1992)
doi:10.1088/0264-9381/9/4/008






\bibitem{Fradkin:1974df}
E.~S.~Fradkin and G.~A.~Vilkovisky,
``S matrix for gravitational field. ii. local measure, general relations, elements of renormalization theory,''
Phys. Rev. D \textbf{8}, 4241-4285 (1973)
doi:10.1103/PhysRevD.8.4241

\bibitem{Donoghue:2020hoh}
J.~F.~Donoghue,
``Cosmological constant and the use of cutoffs,''
Phys. Rev. D \textbf{104}, no.4, 045005 (2021)
doi:10.1103/PhysRevD.104.045005
[arXiv:2009.00728 [hep-th]].

\bibitem{Donoghue:2022chi}
J.~F.~Donoghue,
``Nonlocal partner to the cosmological constant,''
Phys. Rev. D \textbf{105}, no.10, 105025 (2022)
doi:10.1103/PhysRevD.105.105025
[arXiv:2201.12217 [hep-th]].

\bibitem{Donoghue:2014yha}
J.~F.~Donoghue and B.~K.~El-Menoufi,
``Nonlocal quantum effects in cosmology: Quantum memory, nonlocal FLRW equations, and singularity avoidance,''
Phys. Rev. D \textbf{89}, no.10, 104062 (2014)
doi:10.1103/PhysRevD.89.104062
[arXiv:1402.3252 [gr-qc]].

\bibitem{Buchmuller:1988yn}
W.~Buchmuller and N.~Dragon,
``Gauge Fixing and the Cosmological Constant,''
Phys. Lett. B \textbf{223}, 313-317 (1989)
doi:10.1016/0370-2693(89)91608-0
\bibitem{Finkelstein:2000pg}
D.~R.~Finkelstein, A.~A.~Galiautdinov and J.~E.~Baugh,
``Unimodular relativity and cosmological constant,''
J. Math. Phys. \textbf{42}, 340-346 (2001)
doi:10.1063/1.1328077
[arXiv:gr-qc/0009099 [gr-qc]].

\bibitem{Percacci:2017fsy}
R.~Percacci,
``Unimodular quantum gravity and the cosmological constant,''
Found. Phys. \textbf{48}, no.10, 1364-1379 (2018)
doi:10.1007/s10701-018-0189-5
[arXiv:1712.09903 [gr-qc]].


\bibitem{Salvio:2024vfl}
A.~Salvio,
``Unimodular quadratic gravity and the cosmological constant,''
Phys. Lett. B \textbf{856}, 138920 (2024)
doi:10.1016/j.physletb.2024.138920
[arXiv:2406.12958 [hep-th]].

\bibitem{Niedermaier:2006wt}
  M.~Niedermaier and M.~Reuter,
  ``The Asymptotic Safety Scenario in Quantum Gravity,''
  Living Rev.\ Rel.\  {\bf 9}, 5 (2006).
  doi:10.12942/lrr-2006-5

\bibitem{Reuter:1996cp}
  M.~Reuter,
  ``Nonperturbative evolution equation for quantum gravity,''
  Phys.\ Rev.\ D {\bf 57}, 971 (1998)
  doi:10.1103/PhysRevD.57.971
  [hep-th/9605030].

\bibitem{Percacci:2017fkn}
R.~Percacci,
``An Introduction to Covariant Quantum Gravity and Asymptotic Safety,''
World Scientific, 2017,
ISBN 978-981-320-717-2, 978-981-320-719-6
doi:10.1142/10369
\bibitem{Reuter:2019byg}
  M.~Reuter and F.~Saueressig,
  ``{\it Quantum Gravity and the Functional Renormalization Group : The Road towards Asymptotic Safety},'' (Cambridge University Press, Cambridge, 2018).


\bibitem{Buccio:2023lzo}
D.~Buccio, J.~F.~Donoghue and R.~Percacci,
``Amplitudes and renormalization group techniques: A case study,''
Phys. Rev. D \textbf{109}, no.4, 045008 (2024)
doi:10.1103/PhysRevD.109.045008
[arXiv:2307.00055 [hep-th]].

\bibitem{Donoghue:2023yjt}
J.~F.~Donoghue and G.~Menezes,
``Higher Derivative Sigma Models,''
[arXiv:2308.13704 [hep-th]].

\bibitem{Buccio:2024hys}
D.~Buccio, J.~F.~Donoghue, G.~Menezes and R.~Percacci,
``Physical Running of Couplings in Quadratic Gravity,''
Phys. Rev. Lett. \textbf{133}, no.2, 021604 (2024)
doi:10.1103/PhysRevLett.133.021604
[arXiv:2403.02397 [hep-th]].

\bibitem{Buccio:2024omv}
D.~Buccio, L.~Parente and O.~Zanusso,
``Physical Running in Conformal Gravity and Higher Derivative Scalars,''
[arXiv:2410.21475 [hep-th]].

\bibitem{Cracow} 64th Cracow School of Theorhetical Physics, Zakopane June 2024 https://th-www.if.uj.edu.pl/school/2024/


\end{thebibliography}
\end{document}